\documentclass[12pt,preprint]{aastex}

\newcommand{\uprule}{\rule{0pt}{3.5ex}}
\newcommand{\dorule}{\rule[-2ex]{0pt}{3.0ex}}

\slugcomment{ver. 2009.02.17}

\shorttitle{Interstellar extinction in MIR}
\shortauthors{Nishiyama et al.}

\begin{document}

\title{Interstellar Extinction Law toward the Galactic Center I\hspace{-.1em}I\hspace{-.1em}I : \\
$J$, $H$, $K_S$ bands in the 2MASS and the MKO systems, \\
and 3.6, 4.5, 5.8, 8.0 $\mu$m in the {\it Spitzer}/IRAC system}

\author{Shogo Nishiyama\altaffilmark{1}, 
Motohide Tamura\altaffilmark{2},
Hirofumi Hatano\altaffilmark{3}, 
Daisuke Kato\altaffilmark{4},
Toshihiko Tanab$\mathrm{\acute{e}}$\altaffilmark{5},
Koji Sugitani\altaffilmark{6},
and Tetsuya Nagata\altaffilmark{1}
}

\altaffiltext{1}{Department of Astronomy, Kyoto University, 
Kyoto, 606-8502, Japan}

\altaffiltext{2}{National Astronomical Observatory of Japan, 
Mitaka, Tokyo, 181-8588, Japan}

\altaffiltext{3}{Department of Astrophysics, Nagoya University, 
Nagoya, 464-8602, Japan}

\altaffiltext{4}{Department of Astronomy, Graduate School of Science, 
The University of Tokyo, Bunkyo-ku, Tokyo, 113-0033, Japan}

\altaffiltext{5}{Institute of Astronomy, Graduate School of Science, 
The University of Tokyo, Mitaka, Tokyo, 181-0015, Japan}

\altaffiltext{6}{Graduate School of Natural Sciences, Nagoya City University,
Nagoya, 464-8602, Japan}

\begin{abstract}

We have determined interstellar extinction law toward the Galactic center (GC) 
at the wavelength from 1.2 to 8.0 $\mu$m,
using point sources detected in the IRSF/SIRIUS near-infrared survey
and those in the 2MASS and {\it Spitzer}/IRAC/GLIMPSE I\hspace{-.1em}I catalogs.
The central region 
$\mid l \mid \la 3\fdg0$ and $\mid b \mid \la 1\fdg0$
has been surveyed in the $J$, $H$ and $K_S$ bands
with the IRSF telescope and the SIRIUS camera whose filters are similar to 
the Mauna Kea Observatories (MKO) near-infrared photometric system.
Combined with the GLIMPSE I\hspace{-.1em}I point source catalog,
we made $K_S$ versus $K_S - \lambda$ color-magnitude diagrams
where $\lambda =$ 3.6, 4.5, 5.8, and 8.0 $\mu$m.
The $K_S$ magnitudes of bulge red clump stars 
and the $K_S - \lambda$ colors of red giant branches
are used as a tracer of the reddening vector in the color-magnitude diagrams.
From these magnitudes and colors, we have obtained 
the ratios of total to selective extinction $A_{K_S}/E_{K_S-\lambda}$
for the four IRAC bands.  
Combined with $A_{\lambda}/A_{K_S}$ for the $J$ and $H$ bands derived by Nishiyama et al.,
we obtain
$A_J : A_H : A_{K_S} : A_{[3.6]} : A_{[4.5]} : A_{[5.8]} : A_{[8.0]}
= 3.02:1.73:1:0.50:0.39:0.36:0.43$ for the line of sight toward the GC.
This confirms the flattening of the extinction curve at $\lambda \ga 3 \mu$m
from a simple extrapolation of the power-law extinction at shorter wavelengths,
in accordance with recent studies.  
The extinction law in the 2MASS $JHK_S$ bands has also been calculated,
and a good agreement with that in the MKO system is found. 
Thus, it is established that 
the extinction in the wavelength range of $J$, $H$, and $K_S$ 
is well fitted by a power law of steep decrease 
$A_\lambda \propto \lambda^{-2.0}$ toward the GC.  
In nearby molecular clouds and diffuse interstellar medium, 
the lack of reliable measurements of the total to selective extinction ratios
hampers unambiguous determination of the extinction law; 
however, observational results toward these lines of sight
cannot be reconciled with a single extinction law.

\end{abstract}


\section{Introduction}
\label{sec:intro}

The absolute value of interstellar extinction 
to an individual star is difficult to determine, 
and so is its wavelength dependence.  
Toward the Galactic center (GC), however, we can directly derive 
the wavelength dependence of interstellar extinction, 
assuming only that the center of stellar distribution 
in the lines of sight is 
at the same distance from us and 
that the foreground extinction is patchy; 
such a principle has been employed for star clusters and 
is referred to as the cluster method or variable-extinction method \citep{Kre93}.
By plotting the apparent magnitude versus the color excess 
of a group of stars, 
one obtains a straight line with the slope equal to 
the total to selective extinction ratio, 
e.g., $A_{K_S} / E_{H-K_S}$.  
This variable-extinction method was applied 
for the red clump (RC) stars around the GC 
by \citet{Woz96} at $V$ and $I$, and by \citet{Nishi06a,Nishi08}
as the ``RC method'' at $V$, $J$, $H$, and $K_S$.

\citet{Nishi06a} measured the peak position of the RC stars 
in the color-magnitude diagram (CMD) of each small 
($\sim 4\arcmin \times 4\arcmin$) field in their survey area of 
$\mid l \mid \leq 2\degr$ and $\mid b \mid \leq 1\degr$, and 
the reddening and the extinction were derived 
from the relative shifts of the peak color and the peak magnitude, 
respectively.  
Here the most important point is 
that the magnitude of extinction accompanied by 
a certain amount of reddening can be derived exactly, 
as long as no difference in 1) the mean distance, 
2) the mean absolute magnitude, 
and 3) the mean color of the RC population exists 
among the small fields.  
Since the survey is deep enough and the survey area of a few degrees 
(projected area of a few hundred pc at the GC) does not seem to have 
different RC population correlated with the amount of foreground 
extinction, this assumption is reasonable.  
The ratios of total to selective extinction, 
$A_{K_S} / E_{H-K_S}$, 
$A_{K_S} / E_{J-K_S}$, and 
$A_{H} / E_{J-H}$ were in turn used to 
derive the wavelength dependence of extinction 
$A_{J} : A_{H} : A_{K_S}$.  
When approximated as a power law of the wavelength $\lambda$, 
the interstellar extinction $A_\lambda$ 
toward the GC decreases as $\lambda^{-1.99}$ 
in the wavelength range of $J$, $H$, and $K_S$, 
being steeper to the longer wavelength than 
the \citet{RL85} result $\lambda^{-1.6}$.

Direct measurements of {\it extinction} are extremely difficult 
because of the uncertainty of the distance of light sources.
Therefore, 
the ratios of {\it reddening} (ratios of color excesses; 
e.g., $E_{\lambda-V} / E_{B-V}$) 
are generally determined first, 
and then these color excess ratios are transformed 
into the absolute extinction $A_{\lambda}$ 
with some assumptions.  
For this purpose, 
the total to selective extinction ratio 
$A_V / E_{B-V} = R_V$ 
is often derived somehow, 
either by extrapolation of the color excess diagram 
to the longer wavelength or by a certain assumption, 
and then employed 
with the equation 
\begin{eqnarray}
\frac{A_{\lambda}}{E_{B-V}} = 
\frac{E_{\lambda-V}} {E_{B-V}} +R_V.
\label{eq:Alambda1}
\end{eqnarray}
So long as one compares the color excesses, 
which are obtained directly and accurately, 
meaningful comparison is possible for different determinations,
and 
many attempts have been made in this color-color method.  
However, caution must be exercised in the use of absolute extinction 
$A_\lambda$.

\citet{RL85} determined the extinction law toward the GC 
from 1 to 13 $\mu$m 
by the color excess observations of 
five supergiants near the GC and $o$ Sco.  
They first set 
$E_{V-K} / E_{B-V} =2.744$ for all the stars
and assumed $A_V / E_{B-V} = 3.09 \pm 0.3$ 
for the extinction toward the GC.  
The lower limit of total to selective extinction ratio 3.09 was 
determined from the decrease of extinction in the range of $L, M, 8~\mu$m, 
but the observation at these wavelengths was relatively uncertain.  
Longward of 3 $\mu$m, \citet{Lutz96} and \citet{Lutz99} observed 
hydrogen recombination line emission toward the GC with the 
{\it Infrared Space Observatory} SWS instrument 
and found that the extinction curve is much flatter 
than the \citet{RL85} results.  
\citet{Lutz96} compared the observed fluxes and expected fluxes 
predicted for the H{I\hspace{-.1em}I} region in an aperture of 
$14\arcsec \times 20\arcsec$ centered on Sgr A$^*$.  
They confirmed the applicability of the case B conditions from 
the observations of different upper-level lines.  
It should be noted, however, that \citet{Lutz96} 
had to assume a $K$ band extinction of 3.47 mag 
because the ratio of reddening,
not the absolute extinction,
in the hydrogen line strengths were derived 
also in their determination.

The wavelength dependence of interstellar extinction toward the GC, 
therefore, can be summarized as follows.  
The extinction in the near-infrared wavebands $J, H, K_S$ is fitted well 
by a steep power law $\lambda^{-2}$, 
but decreases only slightly beyond $\lambda = 3~\mu$m, 
and then shows a large maximum due to the silicate absorption at 
$\lambda = 9.7~\mu$m.  
The steep power law in the wavelength range $1-2.5~\mu$m and flat curve in 
$3-4~\mu$m is 
consistent with the polarimetry results \citep{Nagata94}, 
which can be regarded, in a sense, as an absolute determination of 
the difference between the extinctions of two orthogonal axes.

In directions of other than the GC, flat extinction dependence 
in the wavelength range of $3 - 8~\mu$m has also been derived.  
\citet{Inde05} measured the mean color excess ratios from 
the color distributions of stars observed 
with the Infrared Array Camera \citep[IRAC;][]{Fazio04} 
on board {\it Spitzer Space Telescope} ({\it SST}), 
along two lines of sight in the Galactic plane.  
Investigations 
toward five nearby star-forming regions \citep{Flaherty07} 
and the star-forming dense core Barnard 59 \citep{Roman07} 
also suggest 
relatively flat extinction curves from 3 to 8 $\mu$m.  
All these studies obtained color ratios, 
either 
$E_{\lambda-K_S} / E_{J-K_S}$ or
$E_{\lambda-K_S} / E_{H-K_S}$, first.  
Then these color excess ratios are transformed 
into the absolute extinction ratio 
(e.g., $A_{\lambda} / A_{K_S}$) 
on the assumption of one ratio 
(e.g., $A_{H} / A_{K_S}$) 
using the equation 
\begin{eqnarray}
\frac{A_{\lambda}}{A_{K_S}} = 
\left( \frac{A_H}{A_{K_S}} - 1 \right) 
\frac{E_{\lambda-K_S}}{E_{H-K_S}} + 1. 
\label{eq:Alambda}
\end{eqnarray}
Assuming their distribution in the Galactic plane, 
\citet{Inde05} fitted the locus of RC stars 
in a color magnitude diagram and derived 
$A_{H} / A_{K_S} = 1.55 \pm 0.1$.  
Then using the equation (\ref{eq:Alambda}), 
\citet{Inde05}, \citet{Roman07}, and \citet{Flaherty07} 
derived the absolute extinction ratios.   
As pointed out by \citet{Inde05}, 
such extinction ratios of $A_{H} / A_{K_S}$ on the assumption 
of the RC star location in the Galactic plane 
might be more accurate than the extrapolation of the 
$E_{\lambda-K_S} / E_{J-K_S}$ 
curve toward the longer wavelength to get the absolute extinction values 
as in equation (\ref{eq:Alambda1}), 
but one should keep in mind that 
the assumption of the one ratio can lead to large errors in $A_\lambda/A_{K_S}$.  
Color excess ratios with the {\it SST}/IRAC bands 
are different between the star forming regions and 
the diffuse interstellar medium, and 
a difference in the extinction law between them was suggested 
\citep{Flaherty07}.

In this paper, we assume that 
the center of distribution 
in the lines of sight is at the same distance from us
for the RC giants and the giants in the upper red giant branch (RGB), 
and determined the total to selective extinction ratios 
$A_{K_S} / E_{K_S-\lambda}$ 
for the IRAC wavebands
while we use the IRSF/SIRIUS 
\citep[similar to the the Mauna Kea Observatories (MKO) near-infrared photometric system;][]{Tok02}
survey results for the $K_S$ band.  
Also, using the RC and upper RGB stars toward the GC, 
the total to selective extinction ratios 
$A_{K_S} / E_{J_{2MASS}-K_S}$, 
$A_{K_S} / E_{H_{2MASS}-K_S}$, and 
$A_{K_S} / E_{K_{2MASS}-K_S}$ 
in the Two Micron All Sky Survey (2MASS) bands were derived, 
and the agreement to the steep extinction law 
determined by \citet{Nishi06a} is examined in the 2MASS bands.  
We compare these results with the previous determination of 
interstellar reddening in different lines of sight, 
and discuss their difference.

This variation of the ``RC method'' is unique and 
different from any previous determinations of mid-infrared extinction 
because it is free from the transformation equations (\ref{eq:Alambda1}) and (\ref{eq:Alambda}).  
It is on the assumptions of 
the same mean distance and magnitude of the RC star population among 
the small fields, 
of the agreement of the central positions of 
the spatial distributions of the RC and RGB stars, 
and that the upper RGB stars have colors varying only with luminosity.  
The distance to the RC stars seem to be rather constant \citep{Nishi05} 
in this range of Galactic longitudes, and 
it seems reasonable to assume that the RC and RGB stars have similar 
spatial distributions.  
The upper RGB has been used for determining the interstellar reddening and 
extinction by many investigators.  
\citet{Frogel99} used RGB stars with the unreddened $K$ 
magnitude range of 8.0 to 12.5.  
They compared the RGB colors of the inner bulge 
($|l| < 4\degr, |b| < 3\degr$) 
with that of the Baade's Window \citep{Tiede95}, 
and derived the interstellar extinction.  
They also found that 
the amplitude of metallicity variations in the inner bulge does not cause 
large RGB slope changes in the $J-K$ versus $K$ CMD.  
\citet{Schultheis99} used calculated isochrones as reference 
in drawing the extinction map of the inner bulge, and 
\citet{Dutra03} again used the Baade's Window data as reference 
in their determining the extinction within $10\degr$ of the GC. 
Note that although we have no reliable reference of 
reddening-free RGB in the {\it SST}/IRAC bands, 
only the relative shifts of magnitude and color 
are needed in the current work.


\section{Observational Data}
\label{sec:obs}

\subsection{IRSF/SIRIUS}

The central region of our Galaxy,
$\mid l \mid \la 3\fdg0$ and $\mid b \mid \la 1\fdg0$ 
(Fig. \ref{fig:survey}),
was observed from 2002 to 2004
using the NIR camera SIRIUS
\citep[Simultaneous Infrared Imager for Unbiased Survey;][]{Nagas99, Nagay03} 
on the 1.4 m telescope IRSF (Infrared Survey Facility).
The SIRIUS camera can provide $J$ (1.25 $\mu$m), $H$ (1.63 $\mu$m),
and $K_S$ (2.14 $\mu$m) images simultaneously,
with a field of view of 7\farcm7 $\times$ 7\farcm7
and a pixel scale of 0\farcs45.

Data reduction was carried out with 
the IRAF (Imaging Reduction and Analysis Facility)\footnote{
IRAF is distributed by the National Optical Astronomy
Observatory, which is operated by the Association of Universities for
Research in Astronomy, Inc., under cooperative agreement with
the National Science Foundation.}
software package.
Images were pre-reduced following the standard procedures
of near-infrared arrays 
(dark frame subtraction, flat fielding, and sky subtraction).
Photometry, including point spread function (PSF) fitting, was carried out 
with the DAOPHOT package \citep{Stetson87}.
We used the DAOFIND task to identify point sources,
and the sources were then input 
for PSF-fitting photometry to the ALLSTAR task.
About 20 sources were used to construct the PSF for each image.

We performed a photometric calibration with 2MASS point source catalog \citep{Skrutskie06}.
When fitted with the Gauss function,
the histograms of the difference of the 2MASS and SIRIUS magnitudes
($m_{\mathrm{2MASS}} - m_{\mathrm{SIRIUS}}$) in the three bands
for more than $10^5$ stars have standard deviations of $\approx 0.03$,
suggesting that the accuracy of the zero-point calibration for each star is about 0.03 mag.
The averages of the 10$\sigma$ limiting magnitudes were
$J=17.1$, $H=16.6$ and $K_S=15.6$.

Indication of the internal reliability of our photometry 
is obtained from 
overlapped regions between adjacent fields.  
In our observations, images were taken under different sky conditions
and at different nights, even in different years.  
The variation in photometry was found
due to the different PSF models and zero-point correction
used for the analysis of each field.
We have thus calculated magnitude differences of the same stars
in the adjacent fields.
When we use stars whose photometric errors calculated with IRAF
are $<0.01$ mag,
the mean and standard deviation of the magnitude difference
are less than 0.01 mag and 0.03$-$0.04 mag, respectively, in the three bands.

\subsection{GLIMPSE I\hspace{-.1em}I}

The Galactic bulge region was observed
as one of the {\it SST} Legacy programs,
Galactic Legacy Infrared Mid-Plane 
Survey Extraordinaire (GLIMPSE) I\hspace{-.1em}I.
The region was imaged with three 1.2 s exposures at each location using the IRAC,
which is a four channel camera operating simultaneously
in wave bands, [3.6], [4.5], [5.8], and [8.0],
centered on 3.6, 4.4, 5.7, and 7.9 $\mu$m, respectively.

Two catalogs are provided by the GLIMPSE I\hspace{-.1em}I project,
one is a highly reliable Point Source Catalog (GLMI\hspace{-.1em}IC),
and the other is a more complete Point Source Archive (GLMI\hspace{-.1em}IA).
In this study, we use the GLMI\hspace{-.1em}IC.
The criteria to be included in the GLMI\hspace{-.1em}IC
are described in \citet{Meade08}:
e.g., detected at least twice in one band,
at least once in an adjacent band.
The 5 $\sigma$ limiting magnitudes of the point sources are approximately
14, 12, 10.5, and 9.0 mag in the
[3.6], [4.5], [5.8], and [8.0] bands, respectively.
Since the central region ( $\mid l \mid \la 1\degr$ and $\mid b \mid \la 0\fdg75$ )
was observed in another program,
the list of sources in this region is not included in the GLMI\hspace{-.1em}IC.


\section{Reduction and Analysis}
\label{sec:RA}

\begin{figure}[h]
 \begin{center}
   \rotatebox{-90}{
  \epsscale{.40}
    \plotone{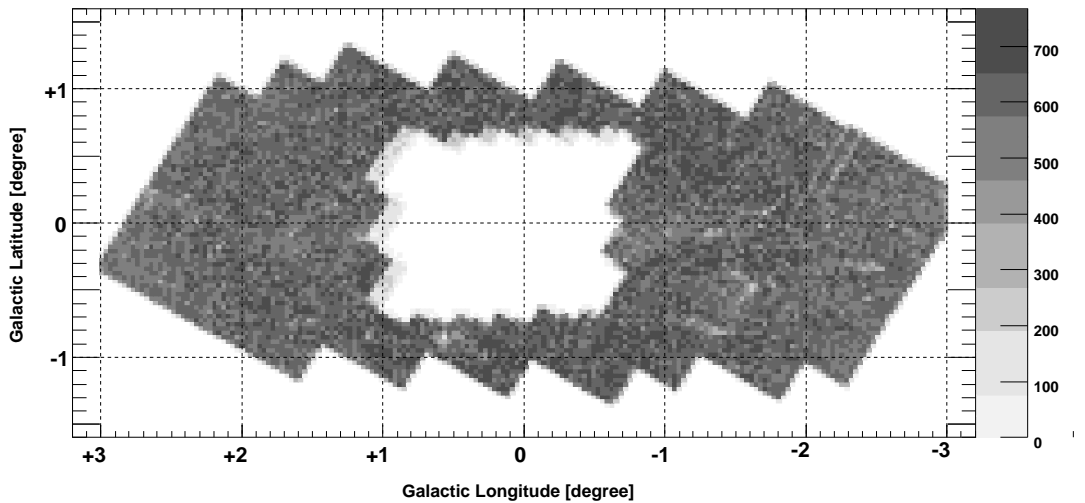}
   }
  \caption{
    A source density map for stars matched between the IRSF/SIRIUS catalog
    and the GLMI\hspace{-.1em}IC.
    The bins are $0\fdg032 \times 0\fdg032$, and
    star count density is given in unit of $[1000 \mathrm{deg}^{-2})]$.
	}
  \label{fig:survey}
 \end{center}
\end{figure}

The stars found with IRSF/SIRIUS and in the GLMI\hspace{-.1em}IC
have been cross-identified
using a simple positional correlation.
The identification was performed with a search radius of 0\farcs6, and
we found $\sim 5.3 \times 10^6$ matches with an rms error 
less than 0\farcs2 in R.A. and Dec,
in the difference between SIRIUS and the GLMI\hspace{-.1em}IC coordinates.
The distribution of matched sources is shown in Fig. \ref{fig:survey}.

\begin{figure}
 \begin{center}
  \epsscale{.60}
    \plotone{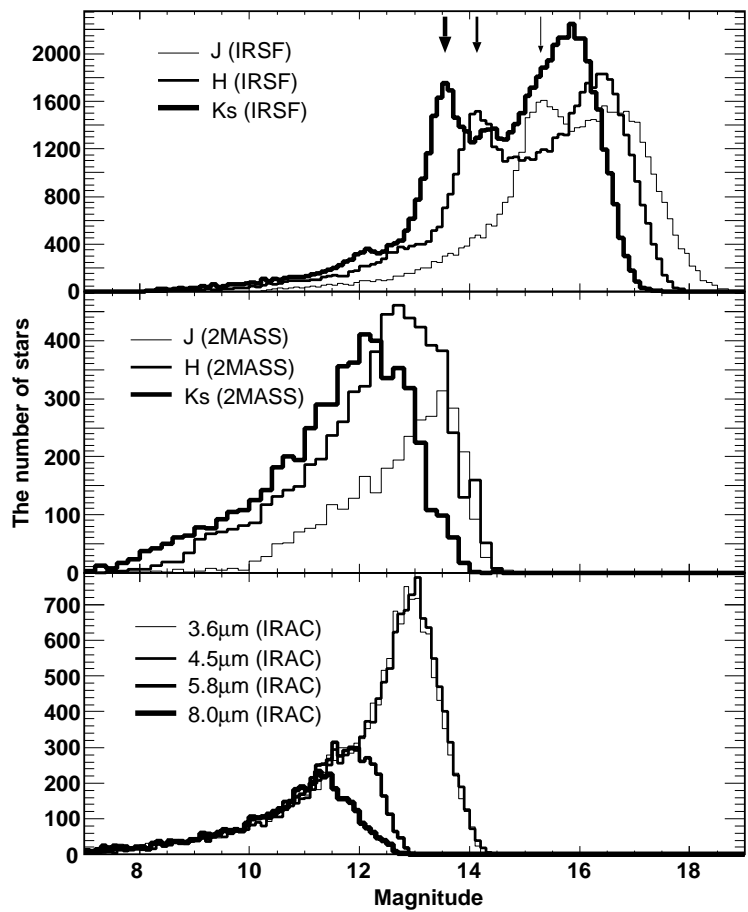}
  \caption{
    Star counts in the $J, H$ and $K_S$ bands
    from IRSF/SIRIUS ({\it top}) and 2MASS ({\it middle}),
    and in the [3.6], [4.5], [5.8], and [8.0] bands
    from {\it SST}/IRAC for one of the fields 
    ($l,b = +0\fdg9, -0\fdg9$) towards the GC.
    Peaks of RC stars can be found only in the top panel,
    and are indicated by arrows.
  }
  \label{fig:LumFAll}
 \end{center}
\end{figure}

Fig. \ref{fig:LumFAll} shows one of the star counts of 
$0\fdg2 \times 0\fdg2$ fields
for IRSF/SIRIUS, 2MASS, and {\it SST}/IRAC.
Clear peaks of RC stars are seen in the star counts of IRSF/SIRIUS
(indicated by arrows),
but not seen in those of 2MASS and {\it SST}/IRAC.
From the limiting magnitudes obtained by GLIMPSE I \citep{Benjamin03},
we expected that RC peaks would be detected at least in the [3.6] and [4.5] bands, 
but source confusion might affect the detection limit of these bands.

To make CMDs involving the {\it SST}/IRAC bands
when we use the ``RC method'' following \citet{Nishi06a}, 
we need to find stars that are detected in the {\it SST}/IRAC bands.  
Here, we use the upper RGB stars to derive the {\it reddening}.
We can use the IRSF/SIRIUS $K_S$ magnitudes of the RC stars for 
the derivation of {\it extinction}. 
Assuming that the RC stars and the upper RGB stars are similarly 
distributed in space with their centers at the GC, 
we will be able to determine in the CMD 
the extinction 
(from RC magnitude shift) 
and the reddening 
(from RGB color shift) of a line of sight toward the GC.  
By plotting the apparent magnitude of the RC stars and 
the color excess of the RGB of each small field, 
we obtain a straight line with the slope equal to 
the total to selective extinction ratio $A_{K_S} / E_{K_S-\lambda}$
for the {\it SST}/IRAC bands.

\begin{figure}
 \begin{minipage}{0.45\linewidth}
   \rotatebox{-90}{
   \epsscale{1.0}
   \plotone{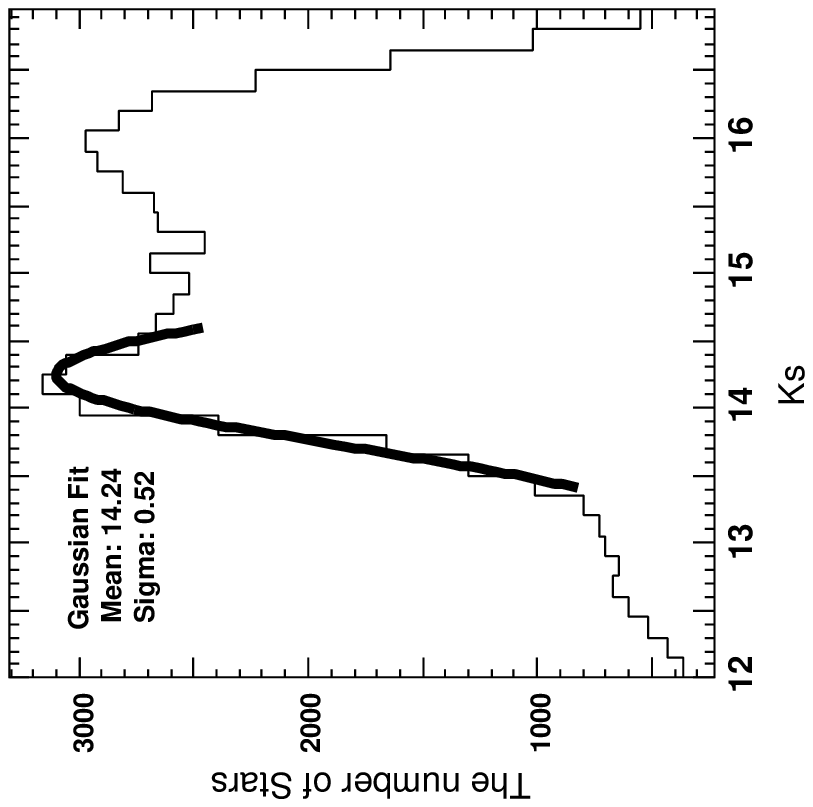}
   }
 \end{minipage}
 \begin{minipage}{0.45\linewidth}
   \hspace{-1.1cm}
   \rotatebox{-90}{
   \epsscale{1.1}
   \plotone{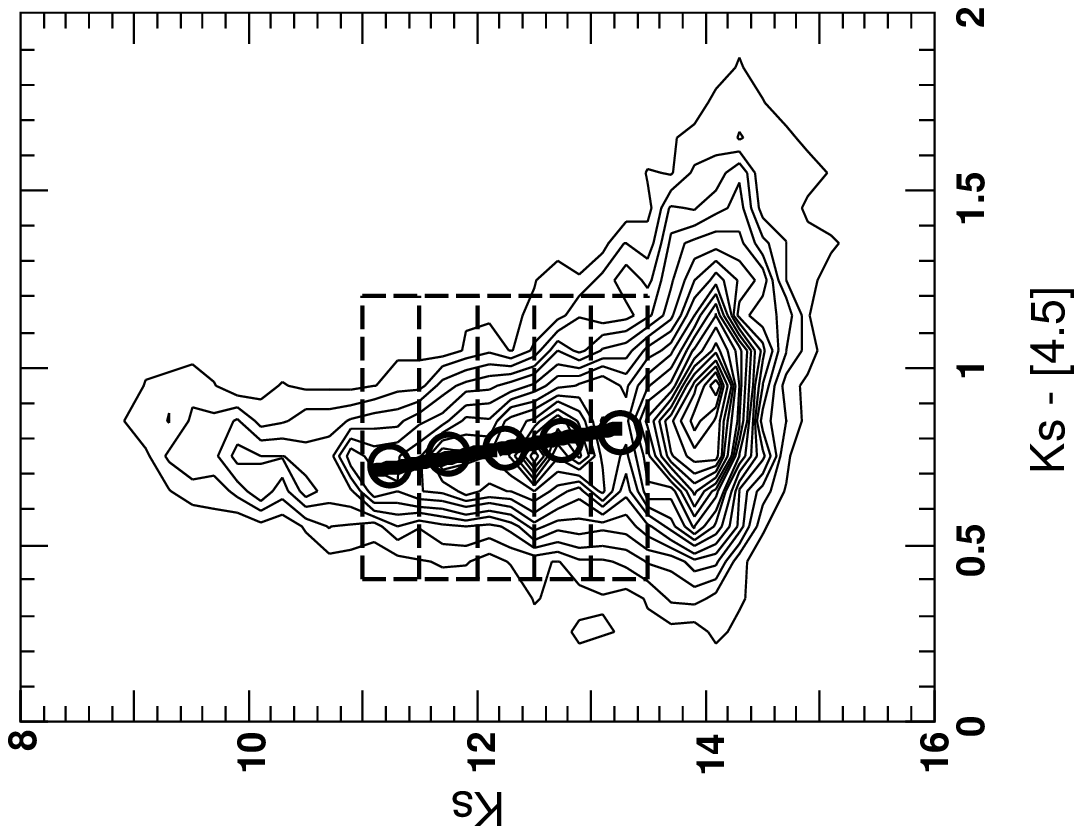}
   }
 \end{minipage}
  \begin{minipage}{0.45\linewidth}
   \begin{center}
   \rotatebox{-90}{
	\epsscale{1.1}
	\plotone{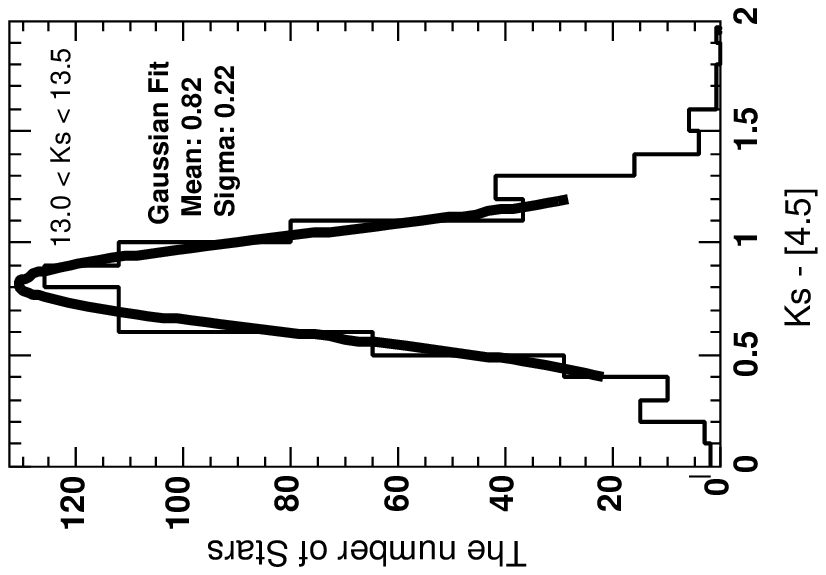}
   }
   \end{center}
  \end{minipage}
  \begin{minipage}{0.45\linewidth}
   \begin{center}
     \epsscale{0.9}
     \plotone{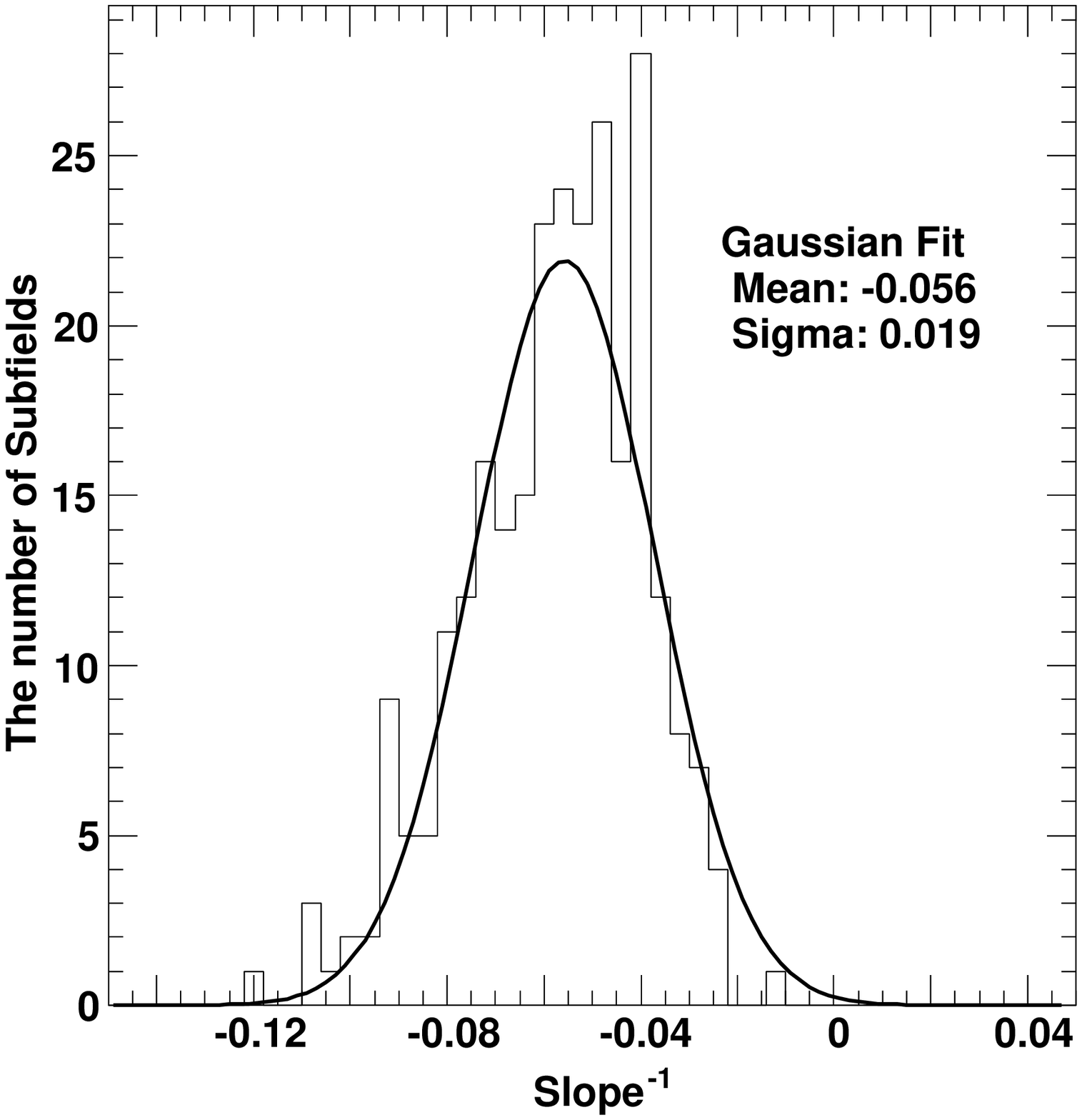}
   \end{center}
  \end{minipage}
   \caption {
     $K_S$-band luminosity function (top left panel)
     and $K_S$ versus $K_S-[4.5]$ CMD (top right panel)
     for one of the fields (centered at $l=-0\fdg3, b=-0\fdg7$).
     The mean $K_S-[4.5]$ colors of stars in the dashed rectangles
     are shown by open circles.
     By fitting the mean colors, we can determine the RGB slope
     for this field.
     Bottom left panel : $K_S-[4.5]$ histograms of stars at $13.0 < K_S < 13.5$,
     which is shown by the bottom rectangle in the top right panel.
     The histogram is fitted with the Gaussian function to obtain the mean of color.
     Bottom right panel : Histogram of (RGB slope)$^{-1}$ 
     in the $K_S$ versus $K_S-[4.5]$ CMDs of the fields.
   }
   \label{fig:CMD}
\end{figure}

As a first step in our analysis,
we divided the survey area (Fig. \ref{fig:survey})
into fields of $0\fdg2 \times 0\fdg2$.
We then made a $K_S$-band luminosity function (LF) for each field.
A sample of the LFs is shown in Fig. \ref{fig:CMD}, top left panel.
A clear peak of RC stars can be found at $K_S \sim 14$ mag.
We determined the $K_S$ peak magnitude (${K_{S}}_{RC}$) by fitting with
the Gaussian function (thick line on the LF in Fig. \ref{fig:CMD}).
\begin{figure}
 \begin{center}
   \rotatebox{180}{
  \epsscale{.80}
    \plotone{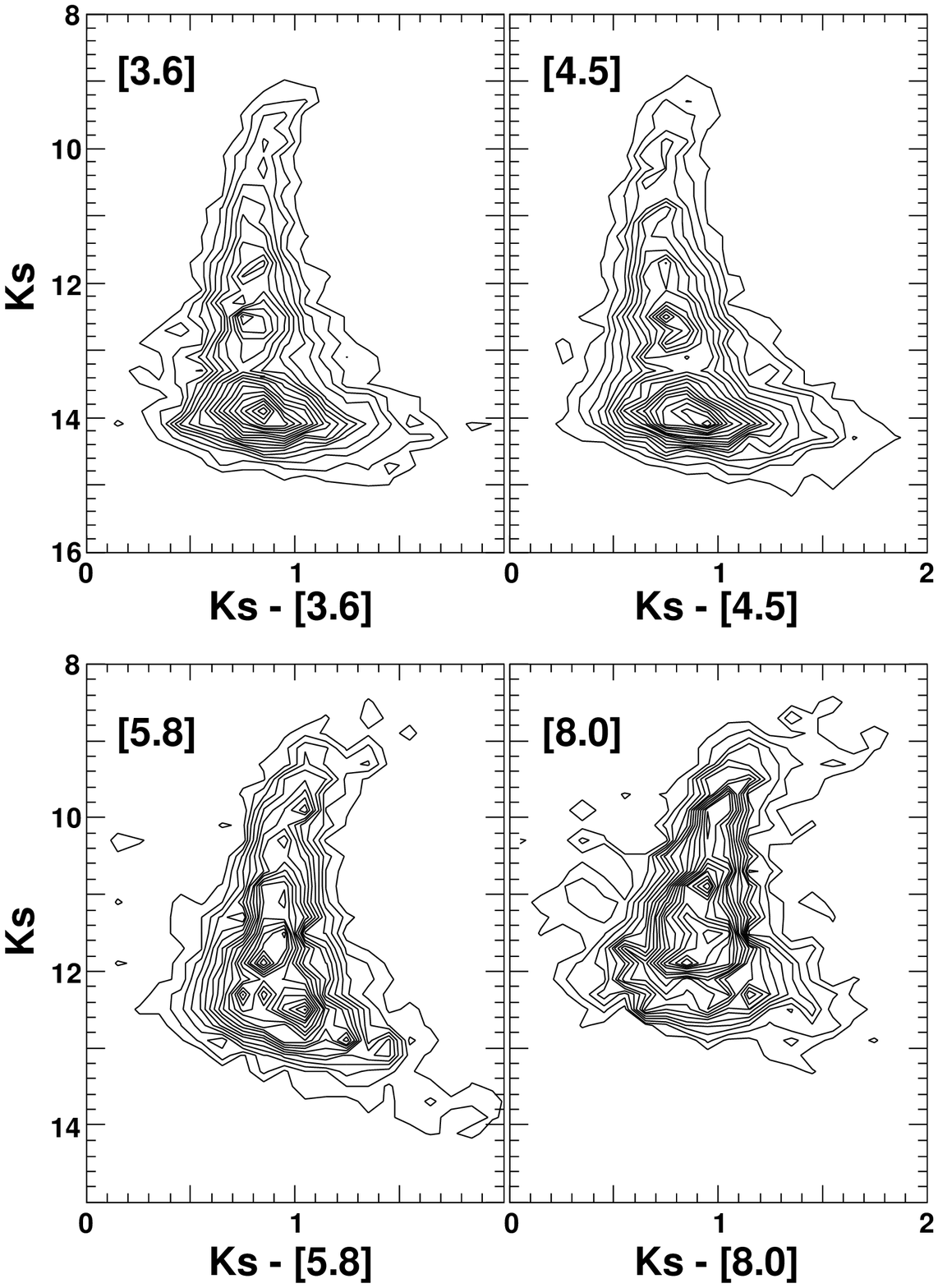}
   }
   \caption{
	 $K_S$ versus $K_S-\lambda$ CMDs
	 of one field (centered at $l=-0\fdg3, b=-0\fdg7$)
	 for $\lambda = [3.6]$ (top left), [4.5] (top right), 
	 [5.8] (bottom left), and [8.0] (bottom right).
  }
  \label{fig:CMDsAll}
 \end{center}
\end{figure}

Second, to determine the color of the RGB,
we made $K_S$ versus $K_S - \lambda$ CMDs 
for each field and IRAC band.
CMDs of one of the fields are shown in Fig. \ref{fig:CMDsAll} 
(see also Fig. \ref{fig:CMD}, top right panel for $\lambda = 4.5~\mu$m).
Third, we determined the RGB slope for each field and 
each IRAC band
on the assumption that the RGB can be fitted by a linear function.
To do this, we divided $K_S$ magnitude of stars on the RGB
into bins of equal (0.5 mag) size,
which is represented in the top right panel, Fig. \ref{fig:CMD}
by dashed rectangles.
Then we calculated the mean of $K_S - \lambda$ color
of stars in each bin, by fitting the color histogram
with the Gaussian function (the bottom left panel in Fig. \ref{fig:CMD}).
The bins near the limiting magnitudes and
those without enough number of stars were excluded.  
The mean $K_S - [4.5]$ colors for the bins are shown by open circles
in the CMD (the top right panel).
The RGB slope for each field was obtained from the least-squares fit 
of the mean $K_S - \lambda$ colors with two free parameters: slope and intercept
(thick line in the top right panel).
The magnitude ranges seem to be wide enough to determine 
the fairly linear part of the RGB, and narrow enough to exclude 
both the brighter asymptotic giant branch stars and 
the fainter RC stars, in these CMDs\footnote{
The fitted magnitude range for the [3.6] and [4.5] bands differs 
slightly from field to field, 
but is approximately 
from $({K_{S}}_{RC}-2.7)$ to $({K_{S}}_{RC}-0.8)$,
where ${K_{S}}_{RC}$ is the peak $K_S$-magnitude of RC stars
for each field, obtained in the first step.
Those for the [5.8] and [8.0] bands are
between $({K_{S}}_{RC}-4.2)$ and $({K_{S}}_{RC}-2.2)$,
and between $({K_{S}}_{RC}-4.2)$ and $({K_{S}}_{RC}-2.6)$.
As shown by the LFs (Fig. \ref{fig:LumFAll}),
the limiting magnitudes in the [5.8] and [8.0] bands are
$\sim 2$ mag brighter than others,
and thus the fitting range is also brighter by $\sim 2$ mag.
}.
We did not find a correlation between the RGB slopes and
extinction for each field.

Fourth, we calculated the mean of the RGB slopes in all the IRAC bands.  
The histogram of the (slopes)$^{-1}$ in 
the $K_S$ versus $K_S - [4.5]$ CMDs 
is shown in the bottom right panel.
The means of the RGB slopes were then obtained by fitting the histograms
with the Gaussian function 
(the bottom right panel in Fig. \ref{fig:CMD}),
as 33.6, 17.9, $-24.6$, and $-19.1$,
for the [3.6], [4.5], [5.8], and [8.0] bands, respectively.

Fifth, the color of the RGB in each field was determined
assuming that the RGBs in all the fields have the same slope as that 
obtained in the third and fourth steps (the mean of the RGB slope).
The mean $K_S - \lambda$ color for each field was determined
by a one-parameter (i.e., intercept) least-squares fit
with the result expressed as the color the RGB stars would have
at the magnitude of the RC stars.
Because of the assumption of constant slope,
what was actually measured is the color at the centroid of 
the RGB color-magnitude distribution.
The coordinate ($K_S - \lambda, {K_{S}}_{RC}$) is the indication
of reddening and extinction of each field
although RC stars are undetected in the GLIMPSE I\hspace{-.1em}I catalog 
(and also in the 2MASS catalog).


\section{Results}
\label{sec:Results}

\begin{figure}
 \begin{center}
   \epsscale{0.80}
  \plotone{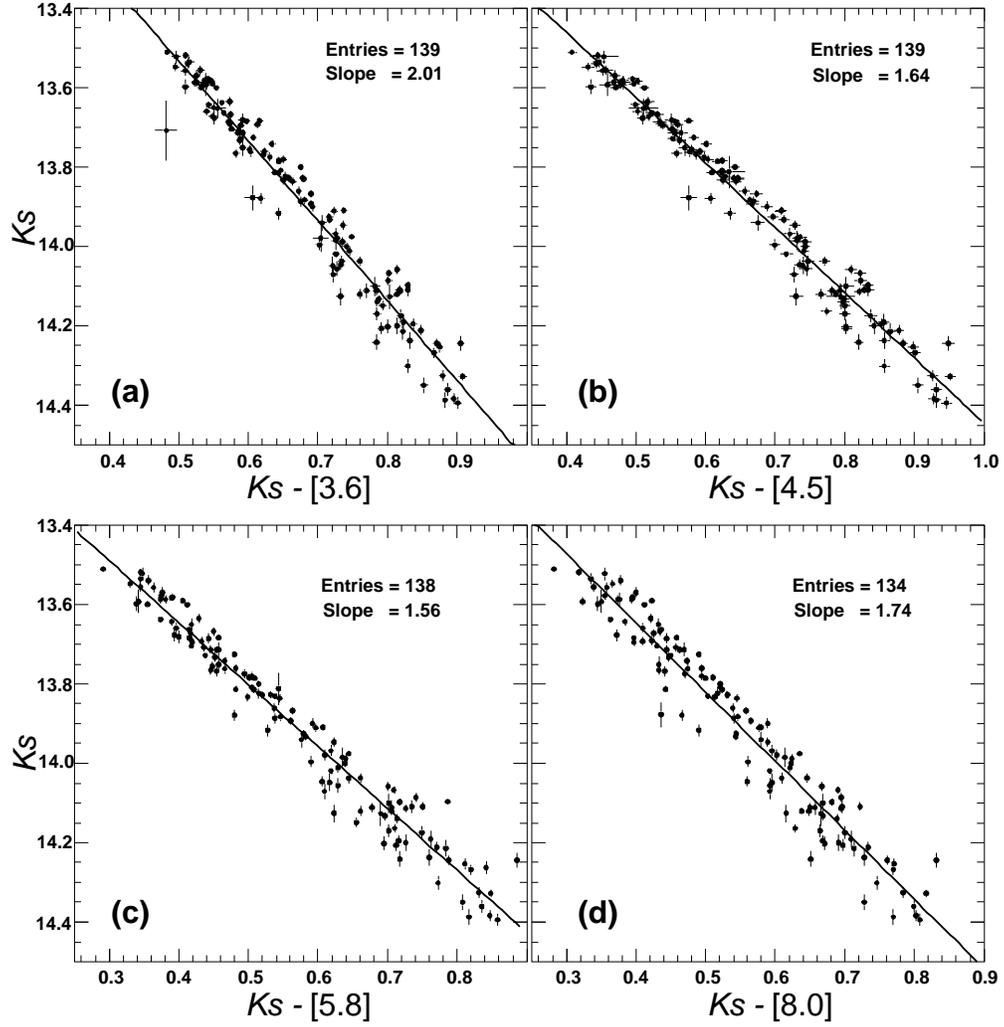}
    \caption{
      (a)-(d): Distribution of ${K_S}_{\mathrm{SIRIUS}}$ magnitudes of a RC peak 
      and ${K_S}_{\mathrm{SIRIUS}} - \lambda$ colors of RGB
      in the $0\fdg2 \times 0\fdg2$ fields
      for the [3.6] (a), [4.5] (b), [5.8] (c), and [8.0] (d) IRAC bands.
    }
    \label{fig:SlIRAC}
 \end{center}
\end{figure}

Fig. \ref{fig:SlIRAC} (a)-(d) shows
the $K_S$ magnitude of the RC peak ${K_{S}}_{RC}$
and the $(K_S - \lambda)$ color of the RGB of each field
in $K_S$ versus $K_S - \lambda$ CMDs.
Error bars on y-axes show uncertainty in RC peak determination,
and those on x-axes come from uncertainty of intercepts
when the least-squares fit is adopted for the mean $K_S - \lambda$ colors.
These errors seem to be underestimated
because the dispersion of the data points is large compared with the error bars.
Hence we estimated the slopes $A_{K_S}/E_{K_S - \lambda}$ and their errors
by applying a least-squares linear fit with $\chi^{2}$ minimization
under the assumption that the errors are all equal.
We show the linear fits applied to the data points in Fig. \ref{fig:SlIRAC},
and the resultant slopes and their errors are listed in Table \ref{tab:slopes}.

\begin{figure}[]
 \begin{center}
   \rotatebox{-90}{
  \epsscale{0.45}
    \plotone{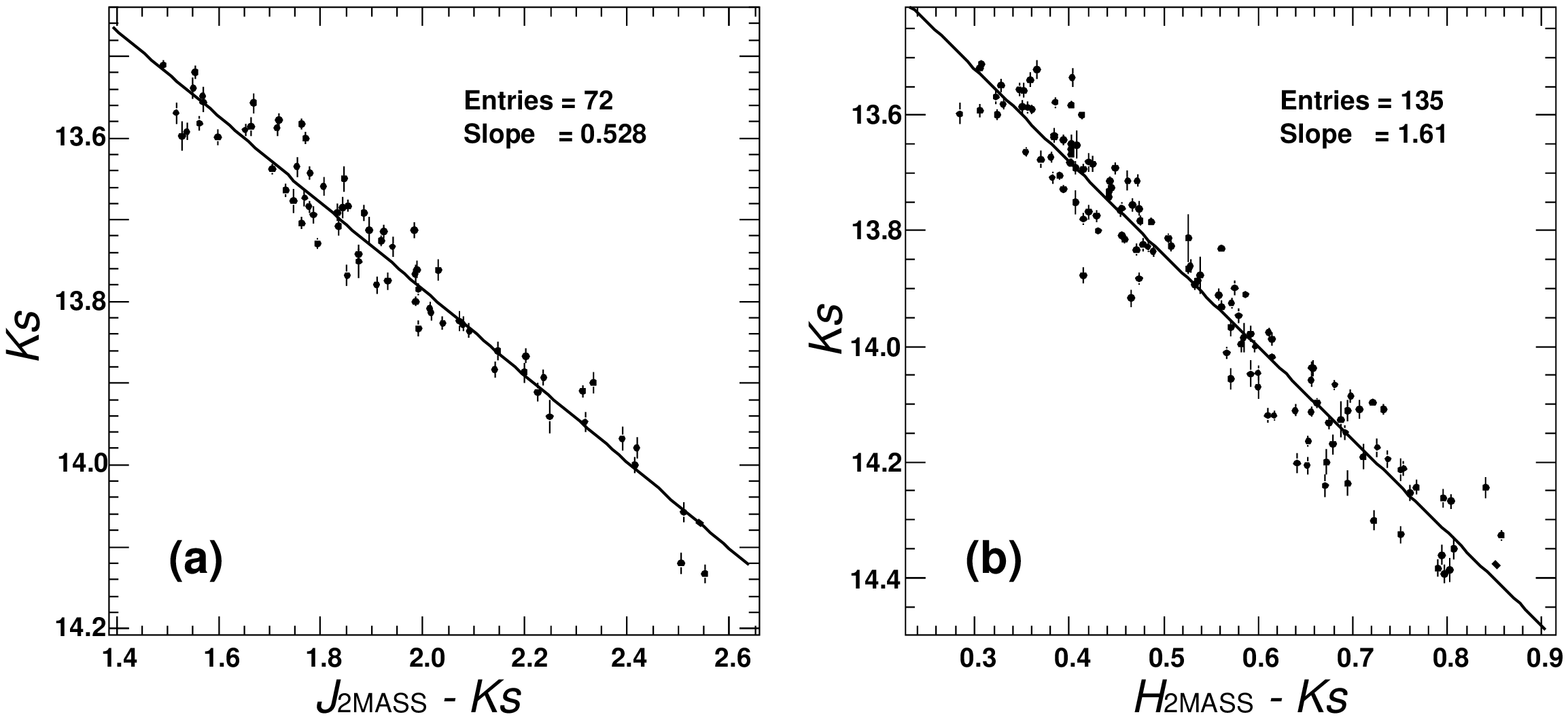}
   }
  \epsscale{0.5}
    \plotone{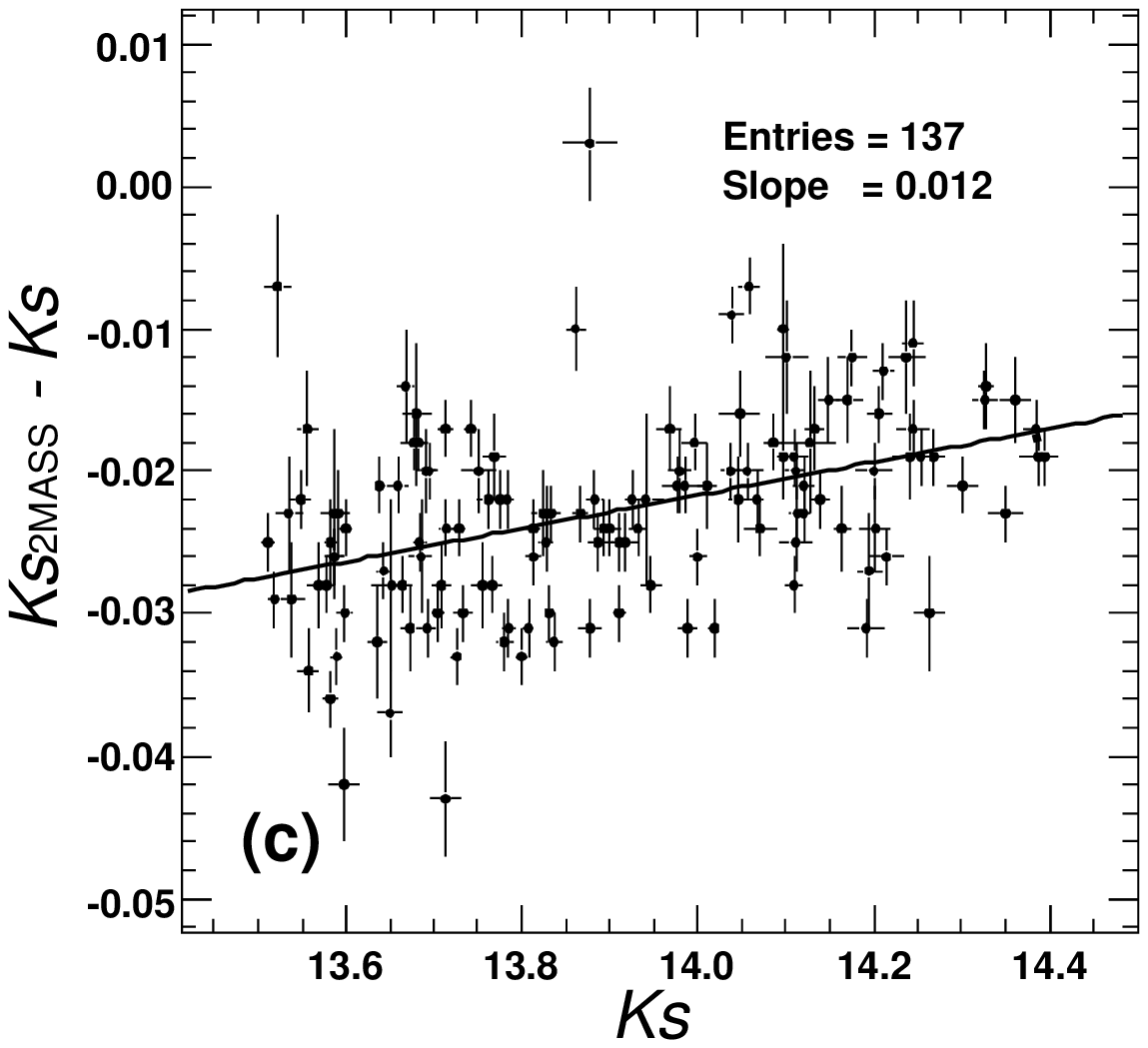}
    \caption{
      (a),(b): Distribution of ${K_S}_{\mathrm{SIRIUS}}$ magnitudes of a RC peak 
      and $\lambda - {K_S}_{\mathrm{SIRIUS}}$ colors of RGB
      in the $0\fdg2 \times 0\fdg2$ fields
      for the 2MASS $J$ (a), and $H$ (b) bands.
      (c): Distribution of ${K_S}_{\mathrm{2MASS}} - {K_S}_{\mathrm{SIRIUS}}$ colors of RGB 
      and ${K_S}_{\mathrm{SIRIUS}}$ magnitudes of a RC peak
      in the $0\fdg2 \times 0\fdg2$ fields.
      To avoid an infinite value of a linear fitting,
      x- and y-axes were interchanged compared to (a) and (b).
    }
    \label{fig:Sl2MASSK}
 \end{center}
\end{figure}

The same procedure described in the previous section was applied 
to 2MASS point sources in the $J$, $H$ and $K_S$ bands,
which are included in the GLMI\hspace{-.1em}IC.
Fig. \ref{fig:Sl2MASSK} show the $K_S$ magnitude of the RC peak 
and the relative ${\lambda}_{\mathrm{2MASS}} - K_S$ color of the RGB 
in $K_S$ versus ${\lambda}_{\mathrm{2MASS}} - K_S$ CMDs.
The dependence of ${K_S}_{\mathrm{2MASS}} - K_S$ on $K_S$ is very small;
hence the same plot in a CMD was made 
for $K_S$ and ${K_S}_{\mathrm{2MASS}}$, 
but the x- and y-axes were interchanged
to avoid an infinite value of the slope (see Fig. \ref{fig:Sl2MASSK}).
The resultant slopes $A_{K_S}/E_{\lambda_{\mathrm{2MASS}}-K_S}$ are 
also summarized in Table \ref{tab:slopes}.  

The uncertainty of the RGB slopes is an error source of this method.
Hence the fifth step described in \S \ref{sec:RA}
was carried out with different RGB slopes,
which are $1 \sigma$ larger and smaller than those previously used.
Here $\sigma$ was obtained when the histograms of (slope)$^{-1}$ are fitted with the Gaussian function.
The changes of $A_{K_S}/E_{K_S - \lambda}$ are only less than a few \% in the IRAC bands,
although those in the 2MASS bands are $2-5$\%.

The ratios of total to selective extinction 
$A_{K_S}/E_{K_S - \lambda} = A_{K_S}/(A_{K_S} - A_{\lambda})$ 
provide us with the ratios of absolute extinction $A_{K_S}: A_{\lambda}$
for the IRAC bands. 
Table \ref{tab:slopes} presents the extinction ratios 
$A_{\lambda}/A_{K_S}$,
which are directly provided by $A_{K_S}/E_{K_S - \lambda}$.
We obtain the wavelength dependence of extinction between $K_S$ and IRAC bands,
$A_{K_S} : A_{[3.6]} : A_{[4.5]} : A_{[5.8]} : A_{[8.0]}
= 1:0.50:0.39:0.36:0.43$.

We also obtain the extinction ratio in the $J$, $H$ and $K_S$ bands
in the 2MASS system,
${A_{J}}_{\mathrm{2MASS}} : {A_{H}}_{\mathrm{2MASS}} : {A_{K_S}}_{\mathrm{2MASS}} = 
2.86\pm0.08 : 1.60\pm0.04 : 1$.
These are slightly smaller than those obtained for the MKO system,
${A_{J}} : {A_{H}} : {A_{K_S}} = 3.02 \pm 0.04 : 1.73 \pm 0.03 : 1$
\citep{Nishi06a},
but the differences of them are within less than
$2 \sigma$ and $3 \sigma$ for
${A_{J}}_{\mathrm{2MASS}}/{A_{K_S}}_{\mathrm{2MASS}}$ and 
${A_{H}}_{\mathrm{2MASS}}/{A_{K_S}}_{\mathrm{2MASS}}$, respectively.

To examine possible variations of the extinction law,
we divided the survey area into quadrants,
N$+$ $(+3\degr > l > 0\degr, +1\degr > b > 0\degr)$,
S$+$ $(+3\degr > l > 0\degr, 0\degr > b > -1\degr)$,
N$-$ $(0\degr > l > -3\degr, +1\degr > b > 0\degr)$, and
S$-$ $(0\degr > l > -3\degr, 0\degr > b > -1\degr)$.
The ratios $A_{K_S}/E_{K_S - \lambda}$ in the quadrants are listed in Table \ref{tab:slopesSubr}.
We do not find significant deviation of the ratios from that for all the data points,
but we find some trends.
The ratios in N$-$ tend to have a smaller value,
while those in S$-$ have a larger one.
These trends, smaller N$-$ and larger S$-$, have been 
obtained in $A_{K_S}/E_{H-K_S}$ \citep{Nishi06a}, and 
similar variations in the extinction law seem to be 
present among the quadrants in the MIR wavebands.  
We note again that 
we do not find statistically significant evidence for differing
the extinction laws in different lines of sight toward the GC
in the IRAC wavebands.

Since most of stars we detected are in the bar structure
whose major axis is oriented at $\sim 20\degr - 40\degr$ with respect to
the Sun-Galactic center line,
different average distance of the giants in a given patch of the sky 
may cause systematic shifts in positions of the RC peaks and the RGB colors
on the $K_S$ versus $K_S-\lambda$ plot. 
As shown in \citet{Nishi05}, mean magnitude of RC stars weakly depends on the Galactic longitude.
However, for changing the ratios of total to selective extinction $A_{K_S}/E_{K_S - \lambda}$,
it is required that the distance to the stars needs to be highly correlated with the reddening.
Such correlation seems to be unlikely to exist\footnote{
We did not find any correlation between extinction values of the fields and their longitude
at $\mid l \mid \leq 2\degr$.
A weak correlation between them at $\mid l \mid > 2\degr$ can be found,
but we confirmed that $A_{K_S}/E_{K_S - \lambda}$ for all the data points
and that for $\mid l \mid \leq 2\degr$ are almost the same.
}.
A good test for the existence of this systematic error is to compare 
the ratios $A_{K_S}/E_{K_S - \lambda}$ for the quadrants,
because the error should be reduced in smaller regions due to smaller difference of the distance.
As described above, 
we found insignificant deviation of $A_{K_S}/E_{K_S - \lambda}$ for the quadrants, 
and very small difference between the ratios for all the survey area and
the weighted mean for the quadrants (see Table \ref{tab:slopesSubr}),
suggesting very small systematic shifts
on the $K_S$ versus $K_S-\lambda$ plots
due to the difference of the distance.

To confirm that the current method using the RGB and RC stars 
in deriving the total to selective extinction ratios 
is consistent with the \citet{Nishi06a} method using only the RC stars, 
we re-analyzed the \citet{Nishi06a} data 
using the RGB and RC stars.  
The resultant ratios are 
$A_{K_S}/E_{H-K_S} = 1.46 \pm 0.03$ and
$A_{K_S}/E_{J-K_S} = 0.499 \pm 0.018$, 
quite consistent with the previously derived ratios 
$1.44 \pm 0.01, 0.494 \pm 0.006$ 
in \citet{Nishi06a}.  
Therefore, the use of RGB colors does not affect the results. 


\begin{table}[h]
 \begin{center}
  \caption{The wavelength dependence of the interstellar extinction in the IRSF/SIRIUS ($J,H,K_S$; similar to the MKO filter system), {\it SST}/IRAC ([3.6],[4.5],[5.8],[8.0]) and 2MASS ($J_{\mathrm{2MASS}}, H_{\mathrm{2MASS}}, {K_S}_{\mathrm{2MASS}}$) bands.}
 \vspace{0.5cm}
  \begin{tabular}[c]{cc|cc}\hline \hline
    \dorule \uprule band & wavelength [$\mu$m] & $A_{K_S}/E_{K_S - \lambda}$\tablenotemark{a} & $A_{\lambda}/A_{K_S}$\tablenotemark{b} \\ \hline
    $J$ & 1.25 \tablenotemark{c} & -- & $3.02 \pm 0.04$ \tablenotemark{e} \\
    $H$ & 1.63 \tablenotemark{c} & -- & $1.73 \pm 0.03$ \tablenotemark{e} \\
    $K_S$& 2.14 \tablenotemark{c} & -- & $1.0$ \\
    $[3.6]$ & 3.545 \tablenotemark{d} & $2.01 \pm 0.04$ & $0.50 \pm 0.01$ \\
    $[4.5]$ & 4.442 \tablenotemark{d} & $1.64 \pm 0.02$ & $0.39 \pm 0.01$ \\
    $[5.8]$ & 5.675 \tablenotemark{d} & $1.56 \pm 0.03$ & $0.36 \pm 0.01$ \\
    $[8.0]$ & 7.760 \tablenotemark{d} & $1.74 \pm 0.04$ & $0.43 \pm 0.01$ \\ \hline
    $J_{\mathrm{2MASS}}$ & 1.240 \tablenotemark{d} & $-0.528 \pm 0.015$ & $2.89 \pm 0.08$ \\
    $H_{\mathrm{2MASS}}$ & 1.664 \tablenotemark{d} & $-1.61 \pm 0.04$ & $1.62 \pm 0.04$ \\
    ${K_S}_{\mathrm{2MASS}}$ & 2.164 \tablenotemark{d} & $-(0.012\pm0.002)^{-1}$ & $1.012 \pm 0.002$ \\ \hline \hline
  \end{tabular}
  \label{tab:slopes}
  \tablerefs{
  \tablenotemark{a} The ratio of total to selective extinction derived from the slopes in Fig. \ref{fig:SlIRAC} and \ref{fig:Sl2MASSK}.
  \tablenotemark{b} The relative extinction to the $K_S$ band (IRSF/SIRIUS system).
  \tablenotemark{c} The mean effective wavelengths of the SIRIUS filters \citep[see \S5.1;][]{Nishi06a}.
  \tablenotemark{d} The isophotal wavelengths of the 2MASS and IRAC filters \citep[see Table 1;][]{Inde05}.
  \tablenotemark{e} From \citet{Nishi06a}.}
 \end{center}
\end{table}


\begin{table}[h]
 \begin{center}
  \caption{The $A_{K_S}/E_{K_S - \lambda}$ ($A_{K_S}/E_{\lambda - K_S}$ for the 2MASS bands) for 
    all the survey area, four quadrants, and their weighted mean}
 \vspace{0.5cm}
  \begin{tabular}[c]{c|c|cccc|c}\hline \hline
    \dorule \uprule band & all data points\tablenotemark{a} & N$+$\tablenotemark{b} & S$+$\tablenotemark{b} & N$-$\tablenotemark{b} & S$-$\tablenotemark{b} & weighted mean\tablenotemark{c} \\ \hline
    $[3.6]$ & $2.01 \pm 0.04$ & $2.01 \pm 0.07$ & $2.03 \pm 0.09$ & $1.85 \pm 0.06$ & $2.05 \pm 0.09$ & $1.96 \pm 0.04$ \\
    $[4.5]$ & $1.64 \pm 0.02$ & $1.62 \pm 0.04$ & $1.60 \pm 0.05$ & $1.59 \pm 0.06$ & $1.69 \pm 0.07$ & $1.62 \pm 0.02$ \\
    $[5.8]$ & $1.56 \pm 0.03$ & $1.56 \pm 0.04$ & $1.52 \pm 0.05$ & $1.44 \pm 0.04$ & $1.64 \pm 0.07$ & $1.52 \pm 0.03$ \\
    $[8.0]$ & $1.74 \pm 0.04$ & $1.82 \pm 0.06$ & $1.67 \pm 0.07$ & $1.80 \pm 0.08$ & $1.83 \pm 0.09$ & $1.77 \pm 0.04$ \\ \hline
    $J_{\mathrm{2M}}$ & $0.528 \pm 0.015$ & $0.497 \pm 0.029$ & $0.534 \pm 0.065$ & $0.494 \pm 0.021$ & $0.587 \pm 0.041$ & $0.510 \pm 0.016$ \\
    $H_{\mathrm{2M}}$ & $1.61 \pm 0.04$ & $1.64 \pm 0.05$ & $1.54 \pm 0.08$ & $1.48 \pm 0.07$ & $1.63 \pm 0.10$ & $1.58 \pm 0.04$ \\ \hline \hline
  \end{tabular}
  \tablerefs{
    \tablenotemark{a} The same value listed in Table \ref{tab:slopes}. 
    \tablenotemark{b} $A_{K_S}/E_{K_S - \lambda}$ for $\sim 3\degr \times \sim 1\fdg5$ quadrants (see \S \ref{sec:Results}).
    \tablenotemark{c} Weighted means of the values for N$+$, S$+$, N$-$, and S$-$.}
  \label{tab:slopesSubr}
 \end{center}
\end{table}


\section{Discussion}
\label{sec:Discuss}

\subsection{Comparison of Wavelength Dependence of Extinction with Previous Studies toward the GC}

\begin{figure}[h]
  \begin{minipage}{0.60\linewidth}
    \begin{center}
      \rotatebox{-90}{
	\epsscale{1.0}
	\plotone{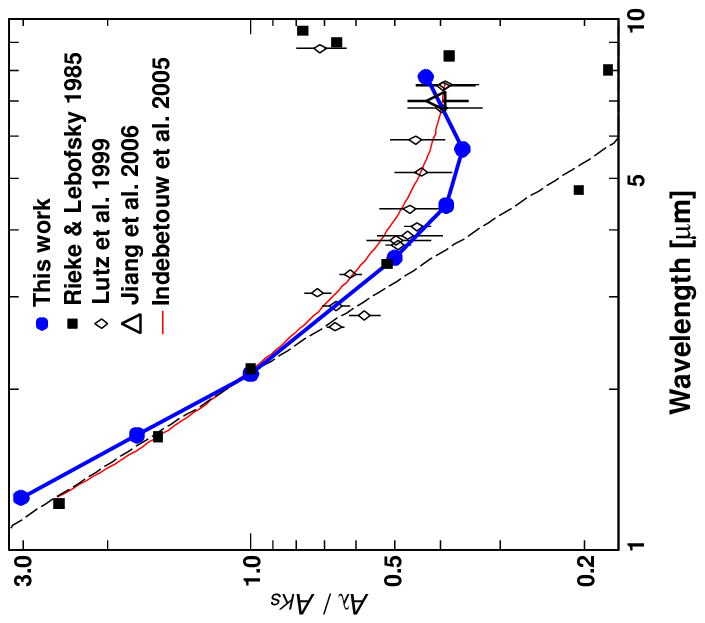}
      }
    \end{center}
  \end{minipage}
  \begin{minipage}{0.35\linewidth}
  \caption{
    The wavelength dependence of extinction ($A_{\lambda}/A_{K_S}$)
    toward the Galactic center region.
    The ratios derived by this study ({\it filled blue circles}),
    \citet[][{\it open diamonds}]{Lutz99}, 
    \citet[][{\it open triangle}]{Jiang06},
    and \citet[][{\it filled squares}]{RL85} are represented.
    Note that in the case of \citet{RL85},
    $A_{\lambda}/A_{K}$ instead of $K_S$ is represented.
    A simple power law, $A_{\lambda} \propto \lambda^{-1.75}$ \citep{Draine89}
    is shown by the dashed line.
    Although it is not toward the Galactic center,
	the extinction curve derived by \citet[][equation 4]{Inde05} is shown by
	{\it thin red smooth line} for reference.
  }
  \label{fig:AlAKOrigGC}
  \end{minipage}
\end{figure}

Fig. \ref{fig:AlAKOrigGC} shows the derived $A_{\lambda}/A_{K_S}$ 
toward the  GC \citep[this study;][]{RL85,Lutz99,Jiang06}.
A simple power law of $A_{\lambda} \propto \lambda^{-1.75}$ 
is also represented.
Before the observations by using ISO/SWS,
the wavelength dependence of interstellar extinction 
in the near- to mid-infrared wavelength range (1 - 6 $\mu$m)
was thought to be represented by a power law 
and to be ``universal'' \citep{Draine89}.  
However, the observation of H{I\hspace{-.1em}I} regions around Sgr A* by using ISO/SWS
shed serious doubt on a simple power law, 
and recent observations toward star forming regions with {\it SST}/IRAC
show clear discontinuity of the power law, 
preferring flatter extinction toward longer wavelengths.

Analysis of the spectrum of the GC obtained with ISO/SWS
revealed an extinction law
characterized by a relatively flat behavior at 3 - 8 $\mu$m \citep{Lutz96}.  
The extinction measurements were improved by 
\citet[][{\it open diamonds} in Fig. \ref{fig:AlAKOrigGC}]{Lutz99},
which reinforce the previous flat extinction.
In Fig. \ref{fig:AlAKOrigGC}, their $A_{\lambda}/A_V$ is converted
to $A_{\lambda}/A_{K_S}$ 
by assuming a $\lambda^{-1.99}$ extinction law \citep{Nishi06a} 
from the $K_S$ wavelength to their $\lambda = 2.625~\mu$m point.  
Fig. \ref{fig:AlAKOrigGC} shows that the extinction law derived by \citet{Lutz99}
is very similar to those presented in this work,
but discrepancy is slightly large in the [4.5] and [5.8] bands.
Some of the discrepancies in $\sim3 - 6~\mu$m wavelength between \citet{Lutz99} and our results
may be explained by large absorption features observed toward the Sgr A*.
\citet{Chiar00} analysed spectra (2.4 - 13 $\mu$m) 
toward the Sgr A* and two sources in the Quintuplet cluster,
and discussed composition of dust along the lines of sight to them.
We can see a deep and broad absorption feature around $\sim 6~\mu$m toward the GC
for which \citet{Chiar00} proposed ice mixtures (H$_2$O:NH$_3$:CO$_2$) and HCOOH.
A deep absorption feature of H$_2$O ice at 3 $\mu$m was also detected.
The same spectrum was used for \citet{Lutz99},
and thus their higher extinction at 3.0392, 3.2970, and 5.9082 $\mu$m
can be attributed to the ice along the line of sight.
In contrast, a relatively shallow and narrow absorption
was observed for the sources in the Quintuplet cluster.
Although it is still unclear how deep and broad the features are
along the other lines of sight in the GC region,
and two of the higher data points (4.3765 and 5.1287 $\mu$m)
cannot be attributed to such absorption features,
if the deep and broad absorption features at $\sim$ 3 and 6 $\mu$m
are characteristic to only Sgr A* and its immediate vicinity,
some of the discrepancies can be naturally explained.

Fig. \ref{fig:AlAKOrigGC} shows that the extinction law 
derived by \citet{RL85} is much smaller 
than \citet{Lutz99} and our results at $> 4~\mu$m, and
the decrease in the 1 - 2.5 $\mu$m range 
is also very different from our results.
\citet{CCM89} 
fit the data of \citet{RL85} with a power law $\lambda^{-1.61}$, and 
it has quite often been referred to as the standard extinction law.  
However, since \citet{RL85} laid great emphasis on 
the determination of $A_V /E_{B-V}$ 
in accordance with the reddening data taken before, 
$J$ magnitudes of $o$ Sco, Cyg OB2 No.12, and 
only two stars near the Galactic center were measured with 
relatively large uncertainties in the color excesses 
mainly due to the assumed errors in intrinsic colors of 
these stars.  
Therefore, the color excesses they determined  
$E_{V-J} / E_{B-V} = 2.19\pm 0.04$ and 
$E_{V-H} / E_{B-V} = 2.55\pm 0.02$ 
on the assumption of $E_{V-K} / E_{B-V} = 2.744$
might be in fact compatible 
with the current results 
if we allow for these uncertainties.

\begin{figure}[h]
  \begin{minipage}{0.60\linewidth}
    \begin{center}
      \rotatebox{0}{
	\epsscale{1.0}
	\rotatebox{-90}{
	\plotone{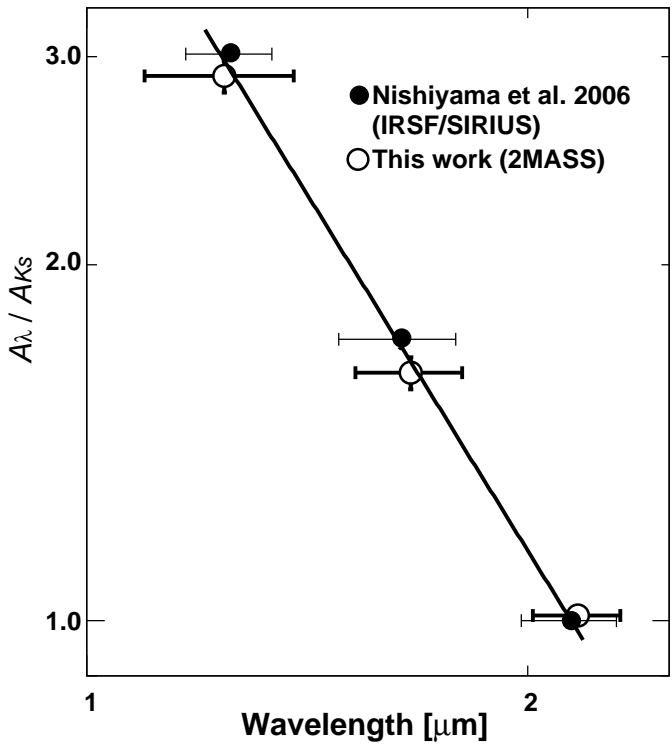}
	}
      }
    \end{center}
  \end{minipage}
  \begin{minipage}{0.35\linewidth}
  \caption{
    The wavelength dependence of extinction ($A_{\lambda}/A_{K_S}$)
    in the $J$, $H$, and $K_S$ bands toward the GC.
    The results of \citet{Nishi06a} 
    in the MKO system are presented by {\it filled circles},
    and that in the 2MASS system by {\it open circles}.  
    A simple power law $A_{\lambda} \propto \lambda^{-2.00}$ 
    fitted to all these points is shown by the straight line.  
    Horizontal bars indicate the filter widths.  
  }
  \label{fig:slopesSI2M}
  \end{minipage}
\end{figure}

The extinction ratios toward the GC in the $J$, $H$ and $K_S$ bands 
in the 2MASS system 
are consistent with the IRSF/SIRIUS results, 
as shown in Table \ref{tab:slopes}.  
These ratios are plotted in Fig. \ref{fig:slopesSI2M}, 
and are consistent with a power law $A_\lambda \propto \lambda^{-2.0}$, 
together with the \citet{Nishi06a} results.  
Therefore, it is established that the interstellar extinction 
toward the GC in the range of $J$, $H$ and $K_S$ bands is 
well fit by a steep power law of $\lambda^{-2.0}$, not $\lambda^{-1.6}$.  
Of course, the power law is only an approximation.  
Since 1) the extinction shows positive deviation from the power law 
as the wavelength increases to $>3~\mu$m as we see above, 
and 2) a number of dust grain models 
\citep[e.g.,][]{Wein01,Dwek04,Voshchinnikov06}
predict such positive deviation, 
it is possible that the power index changes 
in the $J$, $H$, and $K_S$ wavelength range.

\subsection{Comparison in the Ratios of Color Excesses}
\label{subsec:CompRatioCol}

\begin{figure}
 \begin{center}
   \rotatebox{-90}{
     \epsscale{0.7}
     \plotone{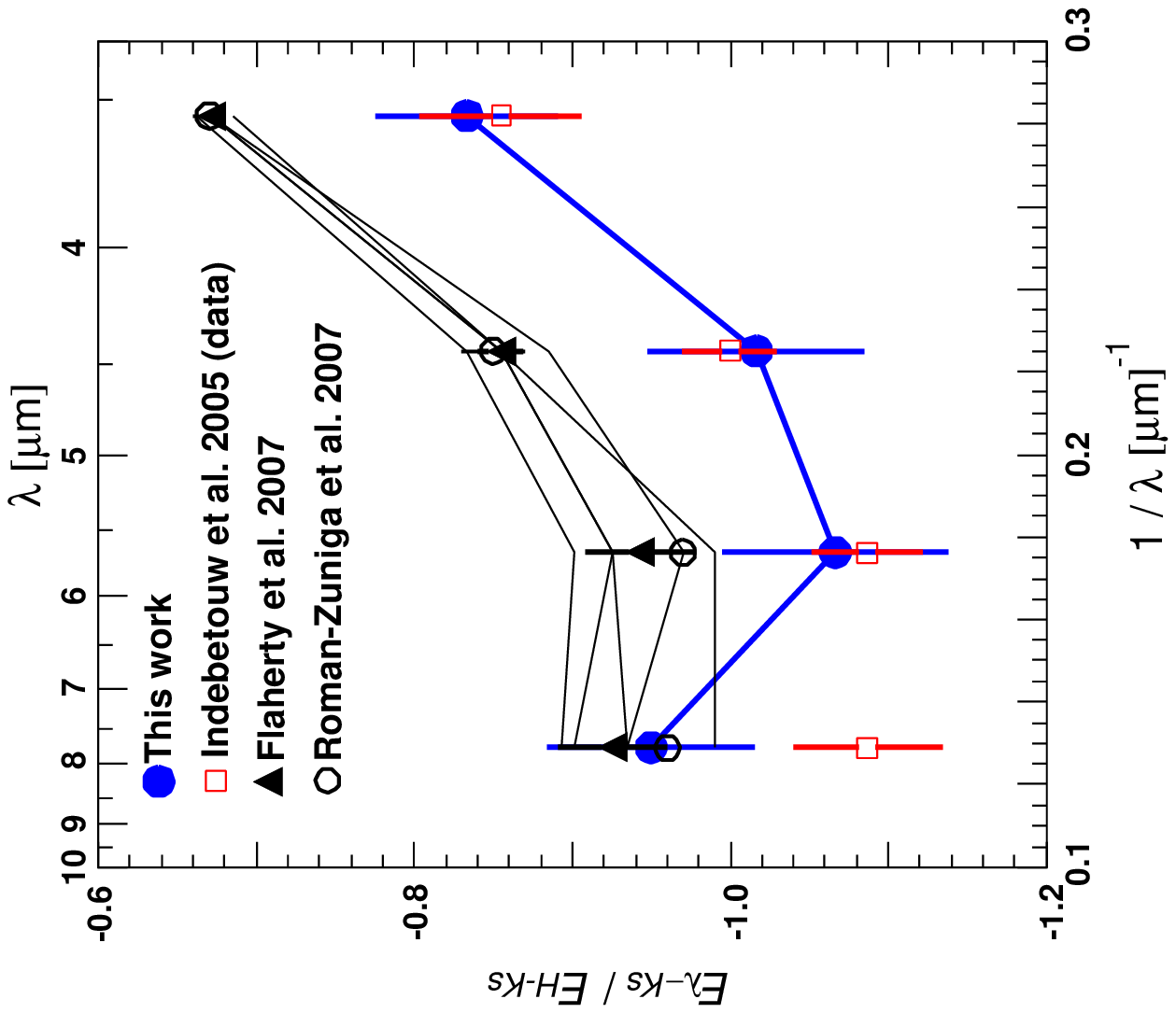}
   }
  \caption{
    Color excess ratio $E_{\lambda-K_S}/E_{H-K_S}$ versus wavelength diagram
    in the IRAC bands
    derived by this study ({\it blue filled circles})
    and \citet[][{\it open circles}]{Roman07}.
    The ratios derived by \citet{Flaherty07} for five star-forming regions ({\it solid lines})
    and their weighted mean ({\it filled triangles}) are also plotted.
    Here all the $H$ and $K_S$ bands are those of the 2MASS survey.  
    The ratio $E_{\lambda-K_S}/E_{H-K_S}$ was not directly measured
    by \citet{Inde05},
    but \citet{Flaherty07} directly determined $E_{\lambda-K_S}/E_{H-K_S}$
    for the same data sets of \citet{Inde05},
    and this is shown by {\it red open squares}.
    So the data points shown in this figure, except ours (blue circles), are
    directly measured from color-color diagrams.
    Only ours are converted from $A_{\lambda}/A_{K_S}$ to $E_{\lambda-K_S}/E_{H-K_S}$.
  }
  \label{fig:ColEHKEKl}
 \end{center}
\end{figure}

A number of authors obtained the ratios of color excesses first, 
and then derived the wavelength dependence 
$A_\lambda / A_{K_S}$ 
of the interstellar extinction.  
However, the assumed ratio $A_H/A_{K_S}$ leads to a 
quite different wavelength dependence, 
as is evident from the equation (\ref{eq:Alambda}) and 
was already pointed out by \citet{Nishi06a} and \citet{Flaherty07}.
Therefore, we convert our 
$A_\lambda / A_{K_S}$ to the ratios of color excesses and compare them 
with the previous results (Fig. \ref{fig:ColEHKEKl}).
This process causes
larger errors 
because of the error propagation, 
but not any uncertainty at all due to assumption of an unknown parameter 
as in the inverse procedure.  

\citet{Flaherty07} and \citet{Roman07} obtained 
$E_{\lambda-K_S} / E_{H-K_S}$
(or its reciprocal)
from the slopes of the distributions of stars in color-color diagrams.
\citet{Inde05} measured the color excess ratios 
$E_{\lambda-K_S} / E_{J-K_S}$, 
but
$E_{\lambda-K_S} / E_{H-K_S}$
was derived from color-color diagrams
({\it open squares})
by \citet[][see their Table 2]{Flaherty07} using the same data sets of 
\citet{Inde05} and the same method
as used in \citet{Flaherty07}.
Thus, direct comparison of the wavelength dependence of extinction 
is possible in color-excess ratios.  
In Fig. \ref{fig:ColEHKEKl}, 
the color-excess ratios 
$E_{\lambda-K_S} / E_{H-K_S}$
are plotted against $\lambda^{-1}$ following the convention 
to compare with scattering calculation.

Differences in 
$E_{\lambda-K_S} / E_{H-K_S}$
seem to exist between the star forming regions and 
the diffuse interstellar medium.  
The ratios toward the star forming dense core \citep{Roman07}
and the nearby star forming regions \citep{Flaherty07} are very similar,
and are in very good agreement within their errors.
As indicated by \citet{Flaherty07},
we can find a systematic separation of the ratio
toward the off-cloud regions \citep{Inde05}
from toward the star forming regions.
The ratios obtained in this study 
(calculated from $A_{\lambda}/A_{K_S}$)
tend to have lower values 
than those for star forming regions, 
in the three bands 
but [8.0].  
The higher extinction in the [8.0] band in the direction of the GC 
than that toward $l = 284\degr$ \citep{Inde05} 
is consistent with the claim that the GC line of sight has 
larger silicate absorption
relative to $A_V$ \citep[larger $\tau_{9.7}/A_{V}$;][]{Roche85}
than other lines of sight \citep[see also Table 1,][]{Draine03}.
A separation between the extinction laws 
for molecular clouds and for the diffuse interstellar medium 
in the color excess ratios $E_{[3.6]-[4.5]} / E_{[4.5]-[5.8]}$ 
suggested by \citet{Flaherty07} is not confirmed 
in Fig. \ref{fig:ColEHKEKl} due to the large uncertainties.  
However, we can find a clear difference in $E_{\lambda-K_S} / E_{H-K_S}$
between the star forming regions and 
the off-cloud regions including toward the GC.

\subsection{Absolute Extinction Ratios in Different Lines of Sight}

Absolute extinction has only been measured for the GC,
and comparison with other lines of sight requires 
an assumption of at least one extinction ratio (e.g., $A_{H} / A_{K_S}$). 
Here, we assume a single power law for the interstellar extinction instead,
and try to derive its index.
\citet{Flaherty07} measured 
$E_{J-K_S} / E_{H-K_S}$ in NGC 2024/2023, NGC 2068/2071, and Serpens, 
which are 
$3.00 \pm 0.04$, $2.98 \pm 0.03$, and $3.05 \pm 0.04$, respectively.  
They also derived the ratio for the $l = 284 \degr$ region 
\citep{Inde05} with the result of $3.07 \pm 0.04$, 
pointing that 
no significant variation seems to exist in the near-infrared 
extinction law through these molecular clouds and 
through the diffuse interstellar medium.  
Assuming the effective wavelengths of the 2MASS observations to be 
1.240, 1.664, and 2.164 $\mu$m \citep{Inde05} and 
that the extinction is approximated by a power law, 
we can derive the power law index for these lines of sight.  
The resultant indices are from $-2.04$ to $-2.18$.  
Thus, the interstellar extinction for all these lines of sight seems to be 
fitted by a single power law of $A_\lambda \propto \lambda^{-2.0}$, 
which is the same as that in the GC, 
very well in the $J$, $H$, and $K_S$ wavelength range.  
However, the extinction law derived by \citet{Inde05} decreases 
much more slowly to the longer wavelength (slower than $\lambda^{-1.7}$), 
and nonetheless their data are shown by \citet{Flaherty07} 
to have a consistent $E_{J-K_S} / E_{H-K_S}$ ratio $3.07 \pm 0.04$.  
(Note that Table 1 in \citet{Inde05} lists smaller $E_{J-K_S} / E_{H-K_S}$ ratios,
possibly because of the difference in the selection criteria of
background stars by them and by \citet{Flaherty07}.)
Therefore, from the color excess ratios one cannot resolve the degeneracy.  
Both a steep power law upto the $K_S$ band similar to the GC and 
a much gentler extinction curve are 
consistent with the color excess ratios. 
In particular, the frequently employed $JHK_S$ color-color diagrams 
can be quite insensitive to the change in extinction law.  
Hence, we cannot determine a value of $A_{H} / A_{K_S}$ 
even from a seemingly similar $E_{J-K_S} / E_{H-K_S}$ values.  
We further should be cautious, 
because line-of-sight differences in $E_{J-K_S} / E_{H-K_S}$ 
do seem to exist.  
For example, \citet{Naoi06} found color excess ratios 
in $\rho$ Oph and Cha I star-forming regions 
consistent with that in the GC, 
but a significantly larger ratio 
in the Coalsack Globule 2 \citep{Naoi07}.

\begin{figure}
 \begin{center}
  \begin{minipage}{0.45\linewidth}
   \rotatebox{-90}{
     \epsscale{1.2}
      \plotone{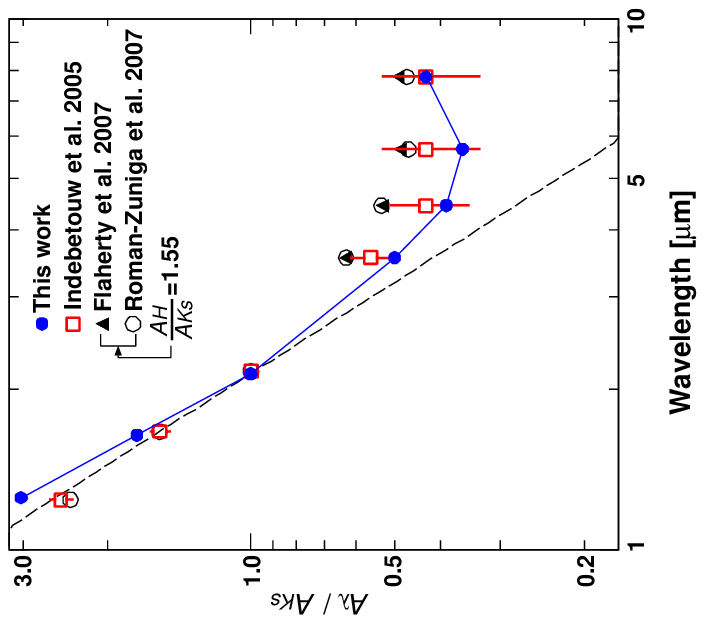}
   }
  \end{minipage}
  \hspace{0.05\linewidth}
  \begin{minipage}{0.45\linewidth}
   \rotatebox{-90}{
     \epsscale{1.2}
     \plotone{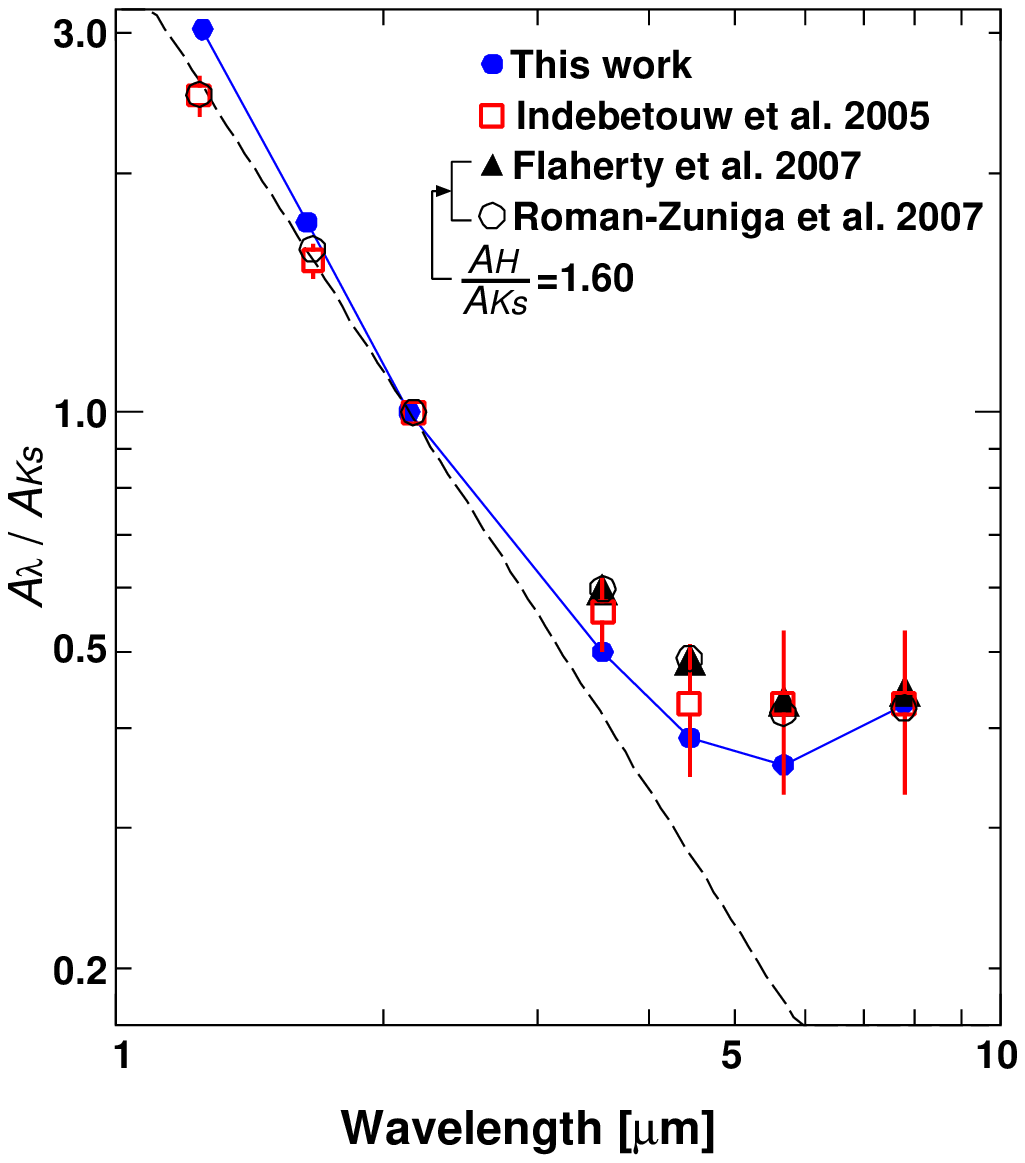}
   }
  \end{minipage}
  \caption{
    Left : The wavelength dependence of extinction ($A_{\lambda}/A_{K_S}$).
    The {\it blue filled circles} (this study) is 
    the extinction law toward the Galactic center,
    and the {\it red open squares} are that for off-cloud regions 
    in the Galactic plane \citep{Inde05}.
    Those for star forming regions derived by 
    \citet{Flaherty07} and \citet{Roman07} are shown by 
    {\it filled triangles} and {\it open circles}, respectively,
    in the assumption of $A_H/A_{K_S} = 1.55$ \citep{Inde05}.
    A simple power law, $A_{\lambda} \propto \lambda^{-1.75}$ is shown 
    by the dashed line \citep{Draine89}.
    Right : The same extinction laws derived by this work and \citet{Inde05}
    are shown, but those for \citet{Flaherty07} and \citet{Roman07} are
    slightly different due to the different assumption of $A_H/A_{K_S} = 1.60$
    corresponding to a steep power law.  
  }
  \label{fig:AlAKOrig}
 \end{center}
\end{figure}

It is quite straightforward 
to determine $A_\lambda / A_{K_S}$ in this work 
from the 
direct measurement of the ratios of total to selective extinction 
$A_{K_S}/E_{K_S-\lambda}$.  
In contrast, 
to derive $A_\lambda / A_{K_S}$ in the equation (2)
in the color-color method is a difficult and crucial task, 
which \citet{Inde05} made from the 
determination of  $A_H/A_{K_S}$.  
Since \citet{Flaherty07} and \citet{Roman07} did not determine $A_H/A_{K_S}$,
they converted their color excess ratios to $A_\lambda / A_{K_S}$
using the ratio $A_H / A_{K_S} = 1.55$ derived by \citet{Inde05}.
However, there is no clear evidence for universality of $A_H/A_{K_S}$,
and a different $A_H/A_{K_S}$ leads to a different $A_\lambda / A_{K_S}$
as described in \S \ref{subsec:CompRatioCol}, 
and as also pointed out by \citet{Flaherty07}.
The difference is shown in Fig. \ref{fig:AlAKOrig}
in which $A_H / A_{K_S} = 1.60$ in addition to $1.55$ is used to 
derive another set of $A_\lambda / A_{K_S}$ ratios 
based on the \citet{Flaherty07} and \citet{Roman07} 
color excess observations.

The ratio of $A_H / A_{K_S} = 1.55$ produces 
larger separation between the line of sights to 
molecular cloud and diffuse interstellar medium.  
The average of the four lines of sight to the star-forming 
regions studied by \citet{Flaherty07} 
and the dark cloud core by \citet{Roman07} 
show higher extinction in the IRAC bands 
than the GC and the $l = 284 \degr$ diffuse medium 
\citep{Inde05}.  
In contrast, if we assume the ratio of $A_H / A_{K_S} = 1.60$, 
the molecular-cloud data points
shift to the GC and the diffuse interstellar medium,
although $A_H / A_{K_S} \approx 1.69$ is needed to make 
the results of \citet{Flaherty07} and \citet{Roman07} 
agree with our results.
Therefore, if we assume that the near-infrared extinction laws are 
different in the dense clouds and diffuse clouds, 
then the difference continues to the mid-infrared, and vice versa.  
Currently, we do not have sufficient data to distinguish 
these two possibilities because 
the measurement of the total to selective extinction is very difficult 
in the lines of sight other than the GC.  
However, the difference in $E_{\lambda-K_S}/E_{H-K_S}$ (see Fig. \ref{fig:ColEHKEKl})
and the required $A_H / A_{K_S}$ of $\approx 1.69$
suggest that all of the observational results discussed in this paper
cannot be reconciled with a single extinction law.

\section{Conclusion}

We have measured the wavelength dependence of interstellar extinction 
toward the GC in the 1.2 - 8.0 $\mu$m region by combining 
the IRSF/SIRIUS infrared observations 
and the 2MASS and {\it SST}/IRAC catalogs.  
The extinction in the wavelength range of $J$, $H$, and $K_S$ 
is well fitted by a power law of steep decrease 
$A_\lambda \propto \lambda^{-2.0}$ toward the GC.  
Furthermore, the flattening of the extinction from a simple extrapolation 
toward the longer wavelength of 
the power law at $\la 3~\mu$m has been confirmed.  
In particular, the extinction has a relatively shallow and broad minimum in 
the {\it SST}/IRAC wavelength range; $A_{[4.5]}$ and $A_{[5.8]}$ are 
only slightly smaller than 0.4 times $A_{{K}_{S}}$.  

This dependence has been derived directly from the observation of 
reddening in proportion to the $K_S$ extinction for the first time 
involving the {\it SST}/IRAC wavebands.  
In this, the $K_S$ magnitudes of RC stars 
and the $K_S - \lambda$ colors of RGB stars serve as a 
tracer of the reddening vector in the color-magnitude diagrams, and 
the ratios of total to selective extinction $A_{K_S}/E_{K_S-\lambda}$ 
have been obtained by a variant of the ``RC method'' originated from 
the variable-extinction method.  
It is interesting to note that this method 
is only sensitive to 
the spatially variable component in the surveyed area; 
when a star cluster suffers patchy $0.8-1.4$mag extinction 
from region to region, then the total to selective extinction ratio 
for the variable $0.6$mag component is derived, but the characteristics of 
the ubiquitous $0.8$ mag extinction component remain unknown.  
This can be its strong point because it is not so sensitive to 
the intrinsic magnitudes and colors of these background stars; 
on the other hand, the location of the dust grains causing 
this variable extinction in the long line of sight 
to the GC is not certain.  
However, since the silicate absorption seems fairly strong, 
the dominant grain population probed in this method is probably 
in the vicinity of the GC.  
It should be also noted that 
the assumptions used here are that the RC stars and RGB stars have 
spatial distributions with their centers in common, and that 
the reddening of RGB stars can be measured precisely from the 
shape of upper giant branch.  

The wavelength dependence seems to be different among the various lines of sight.
In particular, the GC and off-cloud regions show different extinction law from star-forming regions,
although the difference is not very large compared to the observational uncertainties.  
The general behavior of extinction curves, which decreases steeply 
in the near-infrared, but decreases only slightly 
beyond the $K_S$ band, is an important constraint 
when dust grain models are discussed.

\acknowledgements

We thank the staff at the South African Astronomical Observatory (SAAO)
for their support during our observations.
The IRSF/SIRIUS project is supported by Nagoya
University, Kyoto University, and 
the National Astronomical Observatory of Japan
in collaboration with the SAAO.
SN is financially supported by the Japan Society for the Promotion of Science (JSPS) 
through the JSPS Research Fellowship for Young Scientists.
This work was supported by 
Grant-in-Aid for Young Scientists (B) 19740111,
Grant-in-Aid for Scientific Research (A) 19204018, 
Grant-in-Aid for Scientific Research on Priority Area (A) 15071204, 
and Grant-in-Aid for the Global COE Program 
"The Next Generation of Physics, Spun from Universality and Emergence" 
from the Ministry of Education, Culture, Sports, Science and Technology 
(MEXT) of Japan.
This publication makes use of data products from the Two Micron All Sky Survey, 
which is a joint project of the University of Massachusetts and 
the Infrared Processing and Aeronautics and 
Space Administration and the National Science Foundation.


\end{document}